\documentclass[reprint,amsmath,amssymb,aps]{revtex4-1}

\usepackage{graphicx}
\usepackage{bm}
\usepackage{physics}
\usepackage{natbib}
\usepackage{amsfonts,amsthm,bbold}
\usepackage{tikz}
\usepackage[justification=justified]{caption}

\begin{document}

\preprint{APS/123-QED}

\title{Practical Repeaters for Ultra-Long Distance Quantum Communication}

\author{Scott E. Vinay}
 \email{svinay1@sheffield.ac.uk}
\author{Pieter Kok}
 \email{p.kok@sheffield.ac.uk}
\affiliation{ Department of Physics and Astronomy, University of Sheffield}

\date{\today}

\begin{abstract}
\noindent Quantum repeaters enable long-range quantum communication in the presence of attenuation. Here we propose a method to construct a robust quantum repeater network using only existing technology. We combine the ideas of brokered graph-state construction with double-heralded entanglement generation to form a system that is able to perform all parts of the procedure in a way that is highly tolerant to photon loss and imperfections in detectors. We show that when used in quantum key distribution this leads to secure kilohertz bit rates over intercontinental distances.
\end{abstract}

\maketitle

\section{Introduction}
Practical implementations of quantum communication are hampered by the exponential attenuation and decoherence of photons traveling between the two end users, Alice and Bob, putting a maximum limit on the distance over which we can share entangled bits. At present, this limit is on the order of hundreds of kilometers \cite{Tang2016Measurement-Device-IndependentNetwork}. Quantum repeater systems \cite{Briegel1998QuantumCommunication} aim to extend this limit by sharing entangled bits between adjacent stations, and then performing measurements on the qubits within a station to ``distribute'' the entanglement, such that Alice and Bob then share an ideally pure Bell state. This will enable the intercontinental use of new quantum technologies such as absolutely secure encryption  \cite{Scarani2009TheDistribution}, distributed quantum computing \cite{Beals2012EfficientComputing}, teleportation \cite{Bennett1993TeleportingChannels} and more.

Many methods for the construction of fault-tolerant quantum repeaters have been proposed. These include approaches based on measurement-based quantum computation \cite{Zwerger2012Measurement-basedRepeaters}, complex entangled photonic states \cite{Azuma2015All-photonicRepeaters}, quantum error-correction codes on small quantum computers \cite{Jiang2009QuantumEncoding}, advanced multimode memories \cite{Simon2007QuantumMemories} and more \cite{Kok2003ConstructionOptics,Duan2001LongdistanceOptics}. While these are promising methods, many of the ingredients required present formidable experimental challenges. Additionally, there has not necessarily been a suitable answer to the question of how best to generate the initial Bell pairs between the repeaters in a way that retains a high fidelity in situations of non-negligible photon loss and decoherence. This is a crucial element of any proposal for a repeater network, and long distance quantum communication and distributed quantum computing will never be achieved without a satisfactory solution to this problem.

In this letter we address both of these issues by proposing a system based on \emph{doubled-heralded entanglement generation} and \emph{brokered Bell-state measurements}. Critically, these only makes exclusive use of existing technology which has been shown to work reliably in practice, such as in recent demonstrations of Bell's theorem \cite{Hensen2015Loophole-freeKilometres} and teleportation \cite{Pfaff2014UnconditionalQubits}. We describe how the same equipment naturally provides a loss-tolerant way to perform all three parts of the protocol: high-fidelity entanglement generation, loss-tolerant indirect Bell measurements and state distillation. We consider specifically the application of distributing a secret key for secure communication, and an analysis of the relevant errors shows exceptional performance compared to similar protocols. This high performance carries over to other applications which require shared entangled states. We demonstrate this using an in-depth analysis of the errors of the protocol. As such, this work may be constituted as forming a kind of ``threshold theorem," such that if the stated parameters are met, one may be confident that the claimed rates will be practically achievable.

\begin{figure*}[t]
   \centering
     \includegraphics[width=\textwidth]{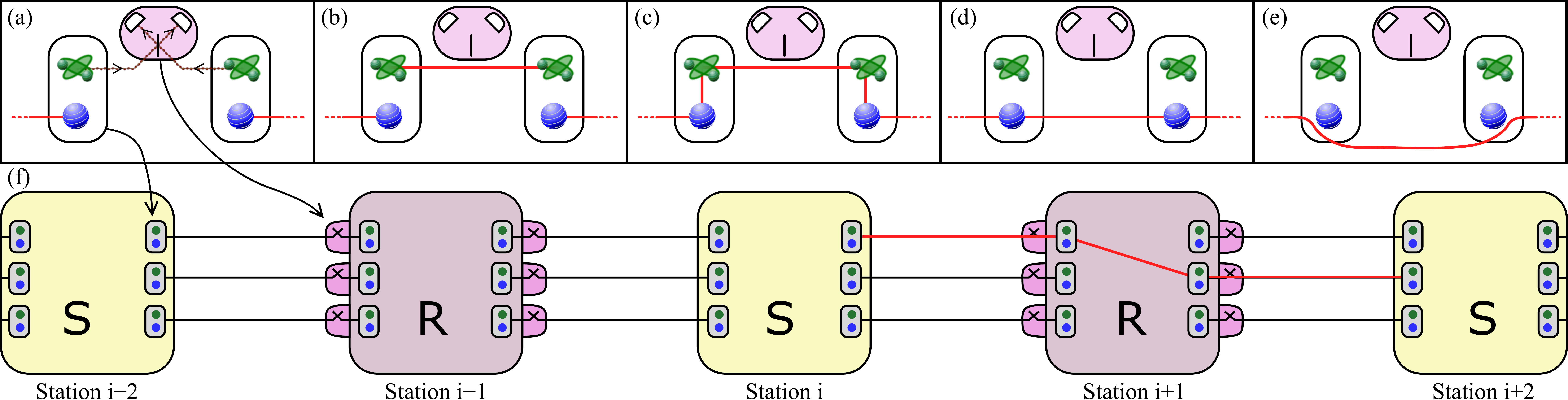}
     \captionsetup{justification=justified}
     \caption{(a)-(e): Illustration of loss-tolerant indirect Bell measurements by brokering. Nuclear spin qubits are in blue and electron spin qubits in green. Red lines represent entanglement connections, which is meant in the sense of an ``edge'' in a graph state. In particular, the electron qubits are connected together by double-heralding in (a)-(b). In (c) these are entangled with the nuclear spins by microwave pulses, and the electron states are measured to give (d). Nuclear spins are then measured to give (e). (f): Four sections of the full repeater protocol, showing S and R type repeater stations. Type S stations send photons to the type R stations, which entangle one qubit from each side. The R stations then send classical signals back to the S stations, which create their own local entanglement. Local measurements then result in long-range Bell pairs between Alice and Bob.}
     \label{fig:proto}
\end{figure*}

\section{Techniques}

Double-heralding \cite{Barrett2005EfficientOptics} is a method by which two distant solid-state qubits may be entangled. It involves the emission of a photon from one of the qubits which is sent to a beam splitter to erase path information. The states of both qubits are transformed by applying a Pauli $\sigma_X$ gate, and the qubits are excited again, possibly leading to another emission. If exactly one photon is detected in each round, we will have projected onto the maximally entangled Bell pair $\frac{1}{\sqrt{2}}\left( \ket{01} + \ket{10} \right)$, where $\ket{0}$ and $\ket{1}$ are the computational basis states of the solid-state qubits. If we fail to detect a photon in either step, both qubits are reinitialized and the process is repeated until a connection is formed. 

The greatest benefit of this method is that, unlike other schemes for remote entanglement generation, the fidelity of the final pair is not affected by attenuation of the photons or imperfections in the detectors. It is also completely unaffected by decoherence of the polarisation or time-bin information of the photon in the optical channel, since we only need to detect the presence of one of more photons. Unlike other proposals for using photons to carry information between distant solid-state qubits, this does not rely on number-resolving detectors. The fidelity will still be affected by dark counts, mode mismatching, and decoherence of the solid-state qubits.

This system is applicable to any physical implementation with two low-lying states and one excited state, but we will consider here specifically the use of nitrogen-vacancy (NV) centers in diamond as has been put forward in other proposals for repeaters \cite{Childress2006Fault-tolerantEmitters,Nemoto2014PhotonicCenters}. This is because NV centers may support two decoupled qubits; one on the spin of the defect electron and one on the spin of one of the nuclear spins. The electrons should be used to establish entanglement connections due to the ease of making measurements of their spin \cite{Jelezko2004ObservationSpin} and their level structure which is appropriate for double-heralding, and the nuclei are used to store entanglement for long time periods, due to their spin coherence times on the order of seconds \cite{Ladd2005CoherenceSilicon}. This may be used for a procedure known as brokering \cite{Benjamin2006BrokeredComputation}.

Brokering, shown in FIG. \ref{fig:proto} (a)-(e), is a procedure whereby two NV centers, A and B, may be entangled without disturbing any existing entanglement that they might have with other centers. This is done by projecting existing entanglement relations onto the nuclear spin qubit. We then try and entangle the electron spin qubits by double-heralding as described above. While a failed attempt at entanglement generation requires us to reset the qubits involved, the fact that the existing entanglement is supported on a separate physical part of the NV center means that the existing entanglement is not disturbed. When the electron-spin qubits of A and B are entangled, a microwave $\pi$ pulse applied to each center applies a controlled-not gate between the electron qubit to the nuclear qubit, entangling them \cite{Jelezko2004ObservationSpin}. Measuring the electron qubits then teleports the entanglement between A and B down on the nuclear spins, and measurement of the nuclear spins removes them from the chain of entanglement, which is equivalent to a Bell state measurement. 

While normal optical Bell state measurements by passive linear gates and no ancilla have a maximum efficiency of 50\%, this procedure has an efficiency limited only by the fidelity of the gates involved and the decoherence times of the nuclear spins. This gate fidelity turns out to be the most important factor in determining the ultimate rate of generation of Bell pairs between Alice and Bob. 

Due to the high fidelity of the Bell pairs that are generated between adjacent stations, this protocol creates high fidelity pairs between Alice and Bob even before any use of distillation. Nevertheless, distillation is a crucial ingredient in extending the reachable range. Here we propose to use the DEJMPS protocol \cite{Deutsch1996QuantumChannels}.

In previous works on repeaters, it was suggested that this protocol may be unsuitable for use in a repeater network, since we require two-way communication to know which attempts have been successful \cite{Nemoto2014PhotonicCenters}. This requires waiting for a time equal to the travel time between distant stations, which we want to avoid since it leads to large decoherences. Alternative suggestions have involved using quantum computers and CSS codes \cite{Li2013LongDistribution,Epping2015GraphNetworks}, but this goes against the philosophy of this work of constructing a simple system which only uses existing technology. The DEJMPS protocol is well suited to our system, since the CNOT gates involved can be implemented by a combination of brokered double-heralding  and local rotations. We can avoid the necessity for long waiting times by implementing \emph{blind DEJMPS}, which is where we assume that all distillation attempts are successful and use the resulting states accordingly. It is only later on after Alice and Bob have measured their qubits that they receive the signals informing them whether the distillation was successful, and hence whether or not they hold a valid key bit. We note that one might still want to use non-blind DEJMPS if gate errors outweigh memory errors.

\section{The Protocol}

We now have the three essential elements to build the repeater network: the creation of long-range Bell pairs, the connection of these pairs within the repeater stations, and the distillation of states, all using the same system of NV centers and microwave pulses. The repeater stations are to be built in two types, type S and type R (for \emph{sender} and \emph{receiver}, shown in FIG \ref{fig:proto}) (f). Each station contains multiple qubits on each side (to connect to the stations before it and after it respectively). The presence of multiple qubits per station decreases the average time that it takes to make at least one entanglement connection between two adjacent repeaters, and so increases the rate of generation of Bell pairs between Alice and Bob even before applying distillation. The full protocol is then implemented as follows.

Type S stations send photons from their qubits to the type R stations before teleporting the state of electron spin qubit onto the nuclear spin. The type R stations use these to try to establish an entanglement connection by double-heralding. Once a type R station has established at least one entanglement connection to the type S stations on each side, it may deterministically entangle them together by using brokering to make a linear graph-like state \cite{Hein2004MultipartyStates}. Classical signals are sent back to the type S repeaters bringing the information of which connections were successful. Once a type S repeater has received such a signal from either side, it may similarly perform a deterministic connection between these NV centers, leaving the final quantum state as a linear chain of entanglement from Alice to Bob via nuclear spins. These nuclear spins may then be removed from the chain by measuring in the computational basis (via projecting back up to the electron spin qubit) leaving Alice and Bob in possession of a pure Bell state. 

In terms of performing the distillation, we should first identify two values, $n_\text{L}$ and $n_\text{S}$, which are the number of sections after which the undistilled fidelity drops to 0.69 and 0.93 respectively. These depend on the inter-repeater distances as well as the error rates in the system. The reasoning behind this is explored in section IV. We attempt to form complete connections over the first $n_\text{L}$ sections so that Alice shares Bell pairs with the $n_\text{L}^{\text{th}}$ station. These pairs should then be distilled to higher fidelity pairs. While the link over the first $n_\text{L}$ stations is being formed, a Bell pair is also formed over the next $n_\text{S}$ stations. This is connected to the distilled pairs over the $n_\text{L}$ sections to form Bell pairs over $n_\text{L} + n_\text{S}$ sections, which are again distilled. We then continue to add Bell pairs over $n_\text{S}$ sections until the Bob is reached. After an agreed-upon length of time, the intermediate stations all measure the state of the nuclear qubits in the computational basis. Alice and Bob then share a high fidelity Bell state.

The setup described above could be used for any of the purposes for which we might want to have long-range entangled states, but we specifically consider here one of the most common: quantum key distribution. This is where Alice and Bob make measurements on their qubits to obtain correlated classical bit strings which can be shown to be completely secure \cite{Gottesman2004SecurityDevices,Scarani2009TheDistribution,Renner2005SecurityDistribution}. This allows Alice and Bob to share an encrypted message of the same length as the secret bit string.

\section{Analysis}

We wish now to derive lower bounds on the secret key rates for both the cases with and without distillation.The main error sources which we identify in affecting the fidelity of the final state are dark counts in the detectors, mismatching the parameters of the NV center cavities, failed gate operations when performing the indirect Bell measurements, and decoherence on the nuclear spins. 
In considering the error analysis we may assume that all measurement results give the $+1$ result, so if all operations are successful Alice and Bob would expect to share $\ket{\Psi^+}\bra{\Psi^+}$ as a final state (measurement results not equal to +1 can be accounted for in classical post-processing). We consider the worst case scenario where a single failed operation maps to the state $\rho = \mathbb{1}_4$. By ``successful operation'' we mean the quantum gates act as expected, the nuclear spins have not decohered, and we have not mistaken a dark count detection as a true detection from a double-heralding round. Let the product of these probabilities be $x$. Their shared state can be described by a Werner state:

\begin{equation}
\rho_W(x) = x \ket{\Psi^+}\bra{\Psi^+} + \frac{1-x}{4} \hspace{1mm} \mathbb{1}_4.
\end{equation}

The quantity that we want to maximise is the secret key rate,

\begin{equation}
K = R\left[ 1 - 2 h_2(e) \right].
\end{equation}

$R$ is the raw rate of bit generation, $e = (1-x)/2$ is the probability of a bit (or, by symmetry, phase) error rate, and $h_2(p) = -p\log(p) - (1-p)\log(1-p)$ is the binary entropy function. The $- 2 h_2(e)$ term represents a fraction of the bits that must be sacrificed to perform error correction and distill the raw key to a secret key \cite{Gottesman2004SecurityDevices}.

In assessing the effects of dark counts, the key parameter of interest is $t_\text{w}$, the \emph{waiting time}. This is the time after the excitation of the electrons in the NV centers that we should wait in order to receive the emitted photons. If this is too small, we will miss the emitted photons, though if it is too great we will certainly measure a dark count, decreasing the fidelity of our states. It should be chosen to maximize $K$.

We have examined the secret key rate for a single elementary section (meaning a single Bell pair established between adjacent stations) against $t_\text{w}$ for different values of detector efficiency and Poissonian dark count rates. We found that choosing $t_\text{w} = 5\tau_\text{q}$ gives a maximal rate for almost any choice of parameters, where $t_\text{q}$ is the timescale over which the excited state of the electron spin decays to emit a photon (see \cite{Barrett2005EfficientOptics}). The robustness of the rate against even the most extreme rates of dark counts indicates that if there is an uncertainty in knowing $\tau_\text{q}$, it is always better to make an over-estimate (and so a larger value for $t_\text{w}$) than an underestimate. We have found that realistic dark count rates affect the fidelity of a single section by less than one part in $10^5$.

In addition to dark counts the main sources of noise are mismatching of modes in adjacent cavities, gate fidelities, and decoherence of the nuclear spins. Mode mismatching has been shown to contribute to an error probability of less than $10^{-3}$ for mismatching either the Jaynes-Cummings constants or the cavity energy constants by up to 5\% \cite{Kok2003ConstructionOptics}. We will include the effects of gate fidelities as a free parameter in our secret key rate, since they simply contribute a constant overhead at each station.  

The error source which required the most consideration is the decoherence of the qubits. This is minimized by utilizing the long-lived nuclear spins, so has little effect on the fidelity for an individual section. The effect becomes pronounced when we consider the full system with $n \gg 1$, where $n$ is the number of sections that are connected together to make the repeater network.

To see how the effects here may be analyzed, consider first the ideal case where every elementary section connects at the same time since the start of the protocol, $t_{\text{avg}} = p_\text{c}^{-1} L_0/c$, where $p_\text{c}$ is the probability for us to make a connection between two adjacent stations in one attempt at double-heralding and $L_0$ is the distance between repeater stations. This is the average time at which a connection between adjacent stations is made. This is given by

\begin{equation}
p_\text{c} = 1 - \left( 1 - \frac{1}{2} e^{-2L_0/L_{\textrm{att}}} \eta^2 \right) ^q,
\end{equation}

\noindent where $L_{\textrm{att}}$ is the attenuation length, $q$ is the number of qubit pairs per station, and $\eta$ is the efficiency of photon emission and collection. We have set it equal to the product of the detector efficiency and the coupling efficiency between the NV center and the optical channel (which may be made deterministic \cite{englund2010deterministic}).

The only decoherence effects here will be a factor of $\exp(-n L_0/c \tau_\text{d})$ contribution to $x$ as the spins decohere slightly while the signals are being sent from the type R stations to the type S stations. This is independent of $p_\text{c}$ since the electron spin qubits are reset for each round of double-heralding. Even for $n=100$ stations at $L_0=25~\text{km}$, $\tau_\text{d} = 1~\text{s}$ this is only a factor of $\sim 1-10^{-5}$ contribution to $x$. Note that we are not considering the contribution of the gate times, since these are mediated by microwave pulses which typically last around 50~ns, compared to the light travel time between stations on the order of tens of microseconds.

A more accurate analysis of the effects of decoherence must take into account the fact that the establishment of Bell pairs across different sections will not all occur at the same time, so the first section to be connected must be kept coherent until the last one has been completed. This is not simply a minor perturbation to the na\"{i}ve situation described in the previous paragraph, since now the non-unit efficiencies of the detectors play a part.

For the set of $n$ sections, let $\{ T_k \}$ be the set of \emph{order statistics}. That is to say, $T_1$ is the time at which the first connection is made, and so on. For an elementary section between two given stations, let $f_t$ be the probability that the connection is formed at a time $t$, and $F_t$ be the probability that it is formed at a time less than or equal to $t$, given by

\begin{equation}
\begin{split}
f_t &= (1-p_\text{c})^{t-1}p_\text{c}, \\
F_t &= 1-(1-p_\text{c})^t.
\end{split}
\end{equation}

The average value of $T_k$ is then given by

\begin{equation}\label{eq:Tk}
\begin{split}
\langle T_k \rangle &= \sum_{t=1}^{\infty} t \sum_{j=0}^{n-k} {{n}\choose{j}} \times  \\ \left[ (1-F_t)^jF_t^{n-j} -\right. &\left.(1-F_t+f_t)^j(F_t-f_t)^{n-j} \right], 
\end{split}
\end{equation}

By taking the worst case scenario that we connect all the odd-numbered sections first (so that we can't make any indirect Bell measurements until as late as possible), we have the following contribution to $x$ from decoherence effects:

\begin{align} \label{eq:xde}
x_{\text{de}} = \exp  \left(-\frac{2L_0}{c\tau_\text{d}}\left[\frac{n}{2} + \langle T_n \rangle + \sum_{k=k_\text{u}}^n \langle T_k \rangle - \sum_{k=1}^{k_\text{l}} \langle T_k \rangle \right]  \right), \vspace{2mm}\\
&  \quad \nonumber
\end{align}

\noindent where $k_\text{u} = \lceil (n+1)/2 + 1 \rceil, k_\text{l} = \lfloor (n+1)/2 \rfloor$, $\lceil \cdot \rceil$ and $\lfloor \cdot \rfloor$ represent the ceiling and floor functions respectively, and $\tau_\text{d}$ is the decoherence timescale ($T_2$ time) of the nuclear spins. The additional factor $n$ comes from the decohering of the nuclear qubits in the time between sending the photons for double heralding and detection

The final ingredient required in finding the overall rate is a decision on when to say that an attempt to make an end-to-end connection has finished, indicating to Alice and Bob that they should then measure their qubits in the $\sigma_Z$ basis to generate a key bit. This may be accomplished by one of two methods:

Method A: When the final section completes, a message is sent from it to Alice and Bob telling them to make the relevant measurements. This will be favorable when $n$ and $\eta$ are both low.

Method B: Decide on a fixed time, $t_\text{f}$ (as a function of $\langle T_n \rangle$), at which Alice and Bob should make their measurements. This will be favorable when $\eta$ is small (so $T_n$ has a narrow distribution) or $n$ is large. With this method, we will ``miss out'' on a fraction $\sum_{t=t_\text{f}}^\infty~P(T_n~=~t)$ of attempted connections.

It has been found that method B, choosing $t_\text{f} = \lceil \langle T_n \rangle + \delta \rceil$ for some buffer value, $\delta$, is better for almost all choices of parameters, with roughly 90\% of connection attempts being successful. However, the behavior of $K$ at a given $\delta$ can be highly erratic with varying $n$, so we should optimize our choice of $\delta$ individually for each choice of parameters (generally on the order of 1 to 5 times $L_0/c$).

Thus we finally arrive at our secret key rate given by Eq \ref{eq:Kmain},
\begin{widetext}
\begin{align} 
K &= \underbrace{ \max_\delta \sum_{t=1}^{\lceil \langle T_n \rangle + \delta \rceil}P(T_n=t)\frac{c }{L_0 \lceil \langle T_n \rangle + \delta \rceil}}_{\text{Raw rate}} \cdot \underbrace{\left[ 1 - 2h_2\left( \frac{1}{2} (1-x_\text{dc}^n x_\text{mm}^n x_\text{ga}^{n-1} x_\text{de}(n)) \right) \right],}_{\text{Correction term}} \label{eq:Kmain}
\end{align}
\end{widetext}

\noindent where $x_\text{dc}$, $x_\text{mm}$, $x_\text{ga}$, and $x_\text{de}$ represent the contribution from dark counts, mode mismatching, Bell measurement gates and nuclear spin decoherence respectively. The raw rate is determined by the light travel time between stations, since (being at the millisecond scale) it is orders of magnitude longer than the timescales involved in referencing the NV centers. 

\subsection*{Distillation}

The DEJMPS procedure of distillation goes as follows. Alice and Bob should share two entangled but noisy pairs of qubits, denoted (A1, B1) and (A2, B2), and locally rotate so that they are diagonal in the Bell basis. Alice and Bob each apply a CNOT gate to their two qubits, with A1 (B1) as the control and A2 (B2) as the target. They then measure A2 (B2) in the computational basis and compare results. If the outcomes match, the control pair is kept for further use and is at a higher fidelity than before. If not, it is discarded. In either case, the target pair is discarded.

When applying DEJMPS to a Werner state, $\rho_\text{W}(x)$, we get the greatest increase in fidelity when $x \approx 0.69$, where fidelity is taken here to mean $\bra{\Psi^+}\rho\ket{\Psi^+}$. As such, we consider a distillation procedure where we perform a distillation operation only at the last repeater station before $x$ is expected to drop below 0.69. Let $n_\text{L}$ be the value of $n$ for which this first happens. The DEJPMS map does not send a Werner state to another Werner state, but instead causes it to tend to a binary mixture of states. Since the calculating the fidelity resulting from repeated application of states becomes analytically intractable, we replace the result of the distillation operation with a Werner state of the same fidelity. This gives an upper bound on the error (and hence a lower bound on the rate) since a Werner state is the highest entropy state of a given fidelity. The probability for the DEJMPS protocol to succeed may be calculated from the diagonal elements of the 2-qubit density matrices as $(\rho_{11}+\rho_{22})^2+(\rho_{33}+\rho_{44})^2$ (where the noisy pairs to be distilled are assumed to be identical). Evaluating for two copies of $\rho_\text{W}(0.69)$, we find that for every 100 noisy pairs we have initially, we expect to keep 37 after a round of distillation.
	
After forming a connection of length $n_\text{L}$, we perform one round of distillation resulting in copies of $\rho_\text{W}(0.74)$. After that we can only afford to connect another $n_\text{S}$ sections at a time before $x$ again drops below 0.69 and we again need to distill (where $n_\text{S}$ may be calculated as the last value of $n$ before which a state of unit fidelity drops to a fidelity of 0.93).

Therefore, for $n\geq n_\text{L}$ (the regime where we intend to start distillation) we get a secret key rate for our protocol of

\begin{equation}
K \geq K_{\text{raw}} 0.37^{\lceil(n-n_\text{L}+1)/n_\text{S}\rceil} \left[ 1-2h_2(0.155) \right],
\end{equation}

\noindent where we are using $\geq$ instead of $=$ since we fix $x$ at the lower bound of 0.69. $K_{\text{raw}}$ is the raw rate term from Eq \ref{eq:Kmain}. Unlike Eq \ref{eq:Kmain} this never drops below zero (since we effectively pin $x$ at 0.69) but at an exponential cost in the raw rate.

We emphasize here that we are considering all noisy pairs to be the same. That is to say, the $k^\text{th}$ order statistic, $T_k$, for any given connection attempt is given by its expectation value. In reality, some connections are going to be established sooner than others and so will have a higher fidelity. There remains the open question of how best to pair up non-identical noisy pairs taken from some distribution. 

\section{Performance}

\begin{figure}[t]
   \centering
     \includegraphics[width=\columnwidth]{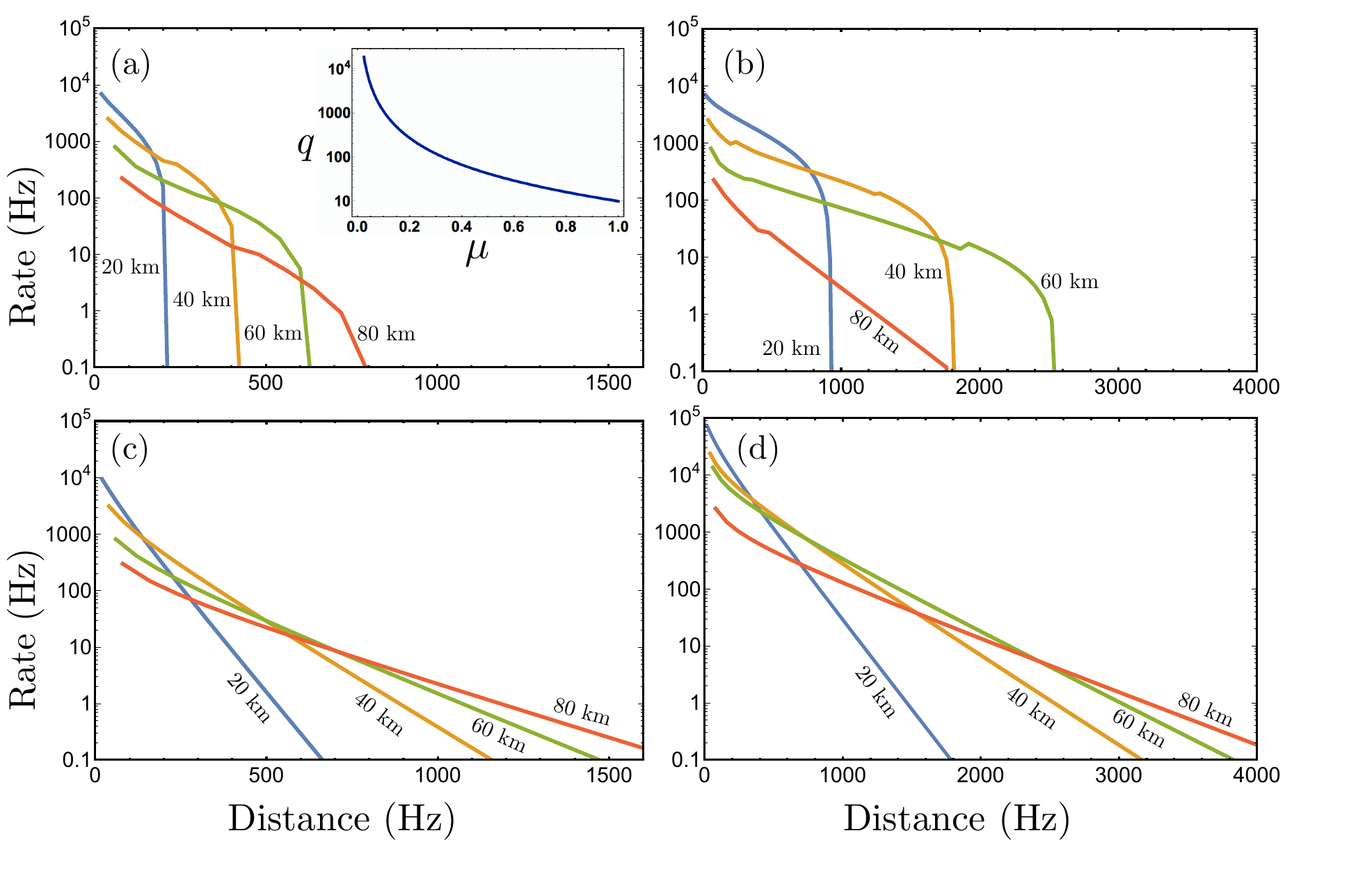}
     \captionsetup{justification=justified}
     \caption{\small Secret key rates are shown for various values of inter-repeater distance, $L_0$ (shown in the values next to the different lines), and quality of brokered Bell measurements, $x_\textrm{ga}$. (a), (c): $x_\textrm{ga} = 0.95$. (b), (d): $x_\textrm{ga} = 0.99$. (a), (b): No distillation. (c), (d): With distillation. Inset: Number of qubit pairs needed to achieve the same rate, $q$, against the proportion of photons on which we post-select, $\mu$.}
     \label{fig:rates}
\end{figure}

In FIG. \ref{fig:rates} we see the rates at which secret keys are generated between Alice and Bob in both the cases of with and without distillation, for an attenuation length of 25~km and detector efficiency of 0.9. We show the rates for varying values of $L_0$ (the distance between adjacent stations) and a fixed 10 pairs of qubits per station. We assume here that we attempt to detect all emitted photons. In various physical implementations this may not be the case, and we may wish to post-select on some fraction $\mu$ of the photons. An example of this may be in NV centers where we wish to use the zero-photon line (since only these photons are perfectly entangled with the emitting center). Recent advancements have produced NV centers with Debye-Waller factors of 0.4, requiring a seven-fold increase in the number of qubits needed per station \cite{johnson2015tunable}. The inset of FIG. \ref{fig:rates} shows the number of qubit pairs needed at each station to achieve the same secret key rates, and how this varies with the fraction upon which we post-select.

\begin{figure}[t]
 \centering
     \includegraphics[width=\columnwidth]{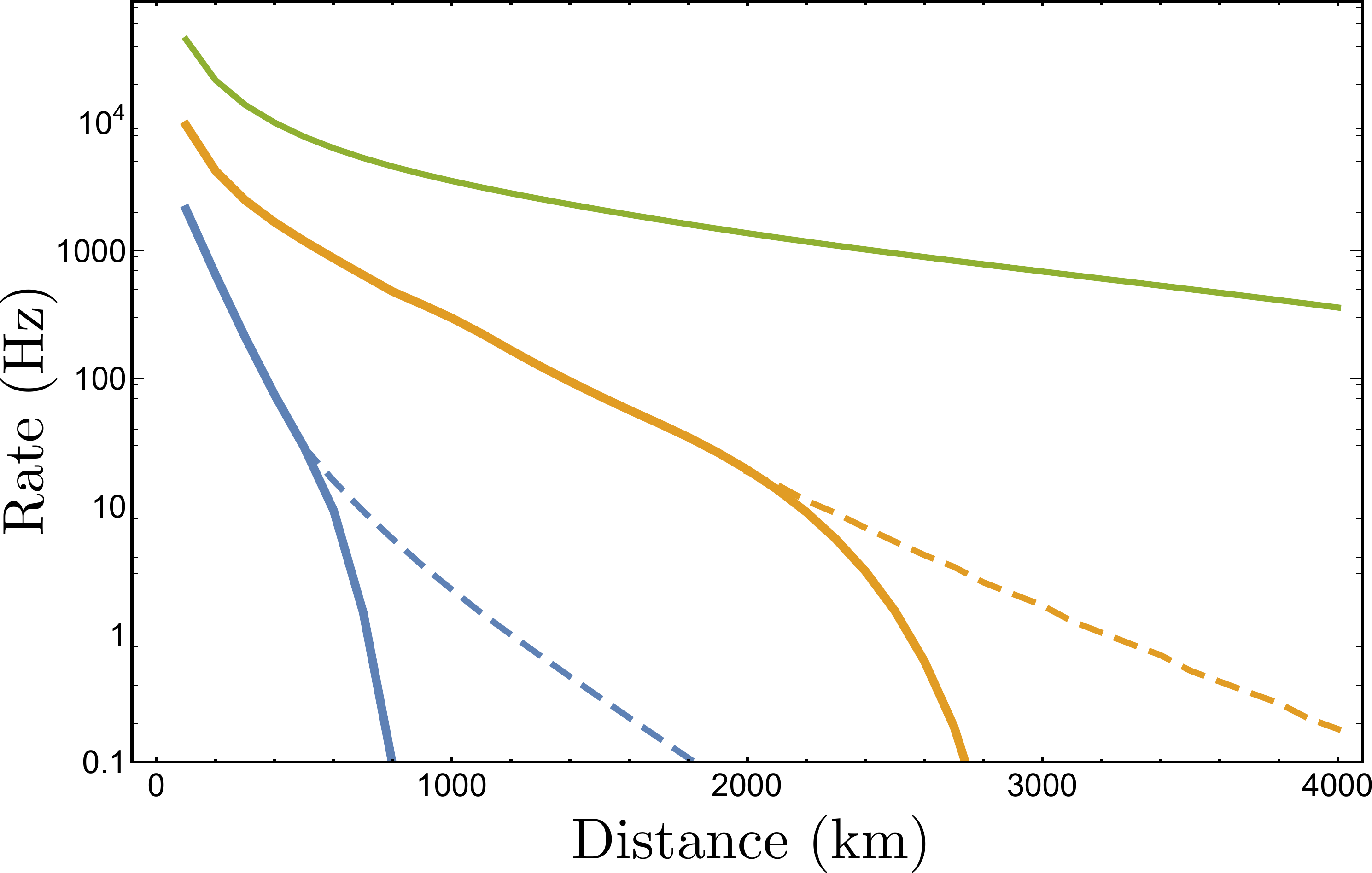}
     \captionsetup{justification=justified}
     \caption{\small Attainable secret key rates for different values of brokering quality, $x_\textrm{ga}$, optimised over values of inter-repeater distance $L_0$. Solid lines are without distillation, dashed lines are with. Top: $x_\textrm{ga} = 0.999$, middle: $x_\textrm{ga} = 0.99$, bottom: $x_\textrm{ga} = 0.95$.}
\label{fig:opt}
\end{figure}

The rates shown here are lower bounds, since we are not including the effects of parallelisation. In reality, when one section forms a connection across one of its pairs of qubits, the others will keep attempting to make connections while waiting for the other sections to connect, meaning the true rate is likely to be far higher. We say that the \emph{gate quality}, $x_\textrm{ga}$, is the probability that no operations involved in carrying out the brokered Bell measurement have suffered heralded or non-heralded failures. We consider both a realistically attainable value of $x_\textrm{ga} = 0.95$ \cite{Leibfried2003ExperimentalGate,Robledo2011HighfidelityRegister} and a reasonable expectation of a future value of $x_\textrm{ga} = 0.99$. For all other parts of the calculation we take worst-case scenarios to ensure that we arrive at a lower bound for the secret key rate. We have fixed physical cost here to only 10 pairs of qubits per station. 
It should be noted that there is not one single choice of $L_0$ that is best for all total distances. For short distances, a smaller $L_0$ gives a higher rate due to the higher rate of connection between adjacent stations, while at higher distances the constant overheads associated with each station (such as the gate errors) begin to dominate the errors, and we get a higher rate by going to a longer $L_0$ and lower number of sections. FIG \ref{fig:opt} shows the optimal choice of $L_0$ for different total lengths, as well as the rate optimized over choices of $L_0$. In addition to the realistic scenarios of $x_\textrm{ga} = 0.95$ and $x_\textrm{ga} = 0.99$, we show the optimized rate in line with the gate quality that is necessary for fault-tolerant quantum computing, $x_\textrm{ga} = 0.999$.

\section{Comparison with other protocols}
We have compared our setup to various others in the existing literature. For a meaningful comparison, it is of course necessary to match the choices of experimental parameters. Since we are comparing against theoretical proposals, there are only a few parameters which are meaningful across all proposals. One such parameter is the local gate efficiency ($x_\textrm{ga}$ in our work), which we have matched with relevant parameters in comparison protocols. Another well-defined value is the resource cost, such as the number of memories, stationary qubits, or subsystems. For a meaningful comparison, we consider here the \emph{normalised} secret key rates, which are the secret key rates divided by the number of qubits used. The third well-defined parameter that we identify is the repetition rate. However, this is included in the overall secret key rate, which we have explained above to be dominated by the light travel time, so does not require additional attention.

One of the closest schemes conceptually is that of Nemoto et. al. \cite{Nemoto2014PhotonicCenters}. This also uses NV centers, but transfers entanglement between the nodes by encoding information in the polarisation of a single photon, which requires the nodes to be equipped with single-photon detectors. A value for local-gate errors of 0.3\% is used, corresponding to a secret key rate of the top line on our Fig. 3. For total distances of 200km and 500km, Nemoto's protocol gives normalised secret key rates of approximately 16 and 3 (where the normalisation divides by the number of qubits used), whilst ours gives normalised rates of 32 and 8. This may seem like a modest improvement, yet we must consider the fact that the two analyses are greatly different. In our protocol we have assumed that the rate of bit errors and phase errors are the same, since the application of indirect Bell measurements to connect two Bell pairs may give any one of the four Bell states as a result, dependent on the outcomes of the measurements. This results in a mixing of phase and bit errors, whereas Nemoto et al. consider phase errors to be dominant.

Additionally our protocol beats other realistic linear-optical repeater schemes such as \cite{Simon2007QuantumMemories,Dur1999QuantumPurification,Krovi2015PracticalSources,piparo2015long} by some orders of magnitude, however gives lower rates than proposals based on advanced encoding schemes \cite{Azuma2015All-photonicRepeaters,Munro2012QuantumMemories}. This is to be expected, since our proposal falls within the category of schemes that are simple to build and do not require large encoded states. In the intermediate regime, there are other protocols. One such is the measurement-based scheme of Ref. \cite{Zwerger2012Measurement-basedRepeaters}, which gives lower normalised rates in the regime of a few thousand kilometers, but has greater reachable distances.

\section{Conclusion}

We have presented a protocol for a quantum repeater network that allows for greater reachable distances and higher secret key rates than other methods in the literature, yet is implementable using today's technology. Unlike most other proposals for such networks, the fidelity of the elementary links is not affected by photon loss, or detectors that do not perfectly count photon number. We have demonstrated that this leads to excellent secret key rates over thousands of kilometers, given sufficiently high gate fidelities. This gives a strong indication that we may be able to have absolutely secure communication over intercontinental distances in the near future.

\subsection*{Acknowledgments}
\noindent We wish to acknowledge the contributions of \mbox{Earl} \mbox{Campbell} for helpful discussions, particularly regarding brokering and NV centers.
\vfill


\begin{thebibliography}{33}%
\makeatletter
\providecommand \@ifxundefined [1]{%
 \@ifx{#1\undefined}
}%
\providecommand \@ifnum [1]{%
 \ifnum #1\expandafter \@firstoftwo
 \else \expandafter \@secondoftwo
 \fi
}%
\providecommand \@ifx [1]{%
 \ifx #1\expandafter \@firstoftwo
 \else \expandafter \@secondoftwo
 \fi
}%
\providecommand \natexlab [1]{#1}%
\providecommand \enquote  [1]{``#1''}%
\providecommand \bibnamefont  [1]{#1}%
\providecommand \bibfnamefont [1]{#1}%
\providecommand \citenamefont [1]{#1}%
\providecommand \href@noop [0]{\@secondoftwo}%
\providecommand \href [0]{\begingroup \@sanitize@url \@href}%
\providecommand \@href[1]{\@@startlink{#1}\@@href}%
\providecommand \@@href[1]{\endgroup#1\@@endlink}%
\providecommand \@sanitize@url [0]{\catcode `\\12\catcode `\$12\catcode
  `\&12\catcode `\#12\catcode `\^12\catcode `\_12\catcode `\%12\relax}%
\providecommand \@@startlink[1]{}%
\providecommand \@@endlink[0]{}%
\providecommand \url  [0]{\begingroup\@sanitize@url \@url }%
\providecommand \@url [1]{\endgroup\@href {#1}{\urlprefix }}%
\providecommand \urlprefix  [0]{URL }%
\providecommand \Eprint [0]{\href }%
\providecommand \doibase [0]{http://dx.doi.org/}%
\providecommand \selectlanguage [0]{\@gobble}%
\providecommand \bibinfo  [0]{\@secondoftwo}%
\providecommand \bibfield  [0]{\@secondoftwo}%
\providecommand \translation [1]{[#1]}%
\providecommand \BibitemOpen [0]{}%
\providecommand \bibitemStop [0]{}%
\providecommand \bibitemNoStop [0]{.\EOS\space}%
\providecommand \EOS [0]{\spacefactor3000\relax}%
\providecommand \BibitemShut  [1]{\csname bibitem#1\endcsname}%
\let\auto@bib@innerbib\@empty
\bibitem [{\citenamefont {Tang}\ \emph {et~al.}(2016)\citenamefont {Tang},
  \citenamefont {Yin}, \citenamefont {Zhao}, \citenamefont {Liu}, \citenamefont
  {Sun}, \citenamefont {Huang}, \citenamefont {Zhang}, \citenamefont {Chen},
  \citenamefont {Zhang}, \citenamefont {You}, \citenamefont {Wang},
  \citenamefont {Liu}, \citenamefont {Lu}, \citenamefont {Jiang}, \citenamefont
  {Ma}, \citenamefont {Zhang}, \citenamefont {Chen},\ and\ \citenamefont
  {Pan}}]{Tang2016Measurement-Device-IndependentNetwork}%
  \BibitemOpen
  \bibfield  {author} {\bibinfo {author} {\bibfnamefont {Y.-L.}\ \bibnamefont
  {Tang}}, \bibinfo {author} {\bibfnamefont {H.-L.}\ \bibnamefont {Yin}},
  \bibinfo {author} {\bibfnamefont {Q.}~\bibnamefont {Zhao}}, \bibinfo {author}
  {\bibfnamefont {H.}~\bibnamefont {Liu}}, \bibinfo {author} {\bibfnamefont
  {X.-X.}\ \bibnamefont {Sun}}, \bibinfo {author} {\bibfnamefont {M.-Q.}\
  \bibnamefont {Huang}}, \bibinfo {author} {\bibfnamefont {W.-J.}\ \bibnamefont
  {Zhang}}, \bibinfo {author} {\bibfnamefont {S.-J.}\ \bibnamefont {Chen}},
  \bibinfo {author} {\bibfnamefont {L.}~\bibnamefont {Zhang}}, \bibinfo
  {author} {\bibfnamefont {L.-X.}\ \bibnamefont {You}}, \bibinfo {author}
  {\bibfnamefont {Z.}~\bibnamefont {Wang}}, \bibinfo {author} {\bibfnamefont
  {Y.}~\bibnamefont {Liu}}, \bibinfo {author} {\bibfnamefont {C.-Y.}\
  \bibnamefont {Lu}}, \bibinfo {author} {\bibfnamefont {X.}~\bibnamefont
  {Jiang}}, \bibinfo {author} {\bibfnamefont {X.}~\bibnamefont {Ma}}, \bibinfo
  {author} {\bibfnamefont {Q.}~\bibnamefont {Zhang}}, \bibinfo {author}
  {\bibfnamefont {T.-Y.}\ \bibnamefont {Chen}}, \ and\ \bibinfo {author}
  {\bibfnamefont {J.-W.}\ \bibnamefont {Pan}},\ }\href {\doibase
  10.1103/PhysRevX.6.011024} {\bibfield  {journal} {\bibinfo  {journal}
  {Physical Review X}\ }\textbf {\bibinfo {volume} {6}},\ \bibinfo {pages}
  {011024} (\bibinfo {year} {2016})}\BibitemShut {NoStop}%
\bibitem [{\citenamefont {Briegel}\ \emph {et~al.}(1998)\citenamefont
  {Briegel}, \citenamefont {D{\"{u}}r}, \citenamefont {Cirac},\ and\
  \citenamefont {Zoller}}]{Briegel1998QuantumCommunication}%
  \BibitemOpen
  \bibfield  {author} {\bibinfo {author} {\bibfnamefont {H.-J.}\ \bibnamefont
  {Briegel}}, \bibinfo {author} {\bibfnamefont {W.}~\bibnamefont {D{\"{u}}r}},
  \bibinfo {author} {\bibfnamefont {J.~I.}\ \bibnamefont {Cirac}}, \ and\
  \bibinfo {author} {\bibfnamefont {P.}~\bibnamefont {Zoller}},\ }\href
  {\doibase 10.1103/PhysRevLett.81.5932} {\bibfield  {journal} {\bibinfo
  {journal} {Physical Review Letters}\ }\textbf {\bibinfo {volume} {81}},\
  \bibinfo {pages} {5932} (\bibinfo {year} {1998})}\BibitemShut {NoStop}%
\bibitem [{\citenamefont {Scarani}\ \emph {et~al.}(2009)\citenamefont
  {Scarani}, \citenamefont {Bechmann-Pasquinucci}, \citenamefont {Cerf},
  \citenamefont {Du\v{s}ek}, \citenamefont {L\"{u}tkenhaus},\ and\
  \citenamefont {Peev}}]{Scarani2009TheDistribution}%
  \BibitemOpen
  \bibfield  {author} {\bibinfo {author} {\bibfnamefont {V.}~\bibnamefont
  {Scarani}}, \bibinfo {author} {\bibfnamefont {H.}~\bibnamefont
  {Bechmann-Pasquinucci}}, \bibinfo {author} {\bibfnamefont {N.~J.}\
  \bibnamefont {Cerf}}, \bibinfo {author} {\bibfnamefont {M.}~\bibnamefont
  {Du\v{s}ek}}, \bibinfo {author} {\bibfnamefont {N.}~\bibnamefont
  {L\"{u}tkenhaus}}, \ and\ \bibinfo {author} {\bibfnamefont {M.}~\bibnamefont
  {Peev}},\ }\href {\doibase 10.1103/RevModPhys.81.1301} {\bibfield  {journal}
  {\bibinfo  {journal} {Reviews of Modern Physics}\ }\textbf {\bibinfo {volume}
  {81}},\ \bibinfo {pages} {1301} (\bibinfo {year} {2009})}\BibitemShut
  {NoStop}%
\bibitem [{\citenamefont {Beals}\ \emph {et~al.}(2013)\citenamefont {Beals},
  \citenamefont {Brierley}, \citenamefont {Gray}, \citenamefont {Harrow},
  \citenamefont {Kutin}, \citenamefont {Linden}, \citenamefont {Shepherd},\
  and\ \citenamefont {Stather}}]{Beals2012EfficientComputing}%
  \BibitemOpen
  \bibfield  {author} {\bibinfo {author} {\bibfnamefont {R.}~\bibnamefont
  {Beals}}, \bibinfo {author} {\bibfnamefont {S.}~\bibnamefont {Brierley}},
  \bibinfo {author} {\bibfnamefont {O.}~\bibnamefont {Gray}}, \bibinfo {author}
  {\bibfnamefont {A.}~\bibnamefont {Harrow}}, \bibinfo {author} {\bibfnamefont
  {S.}~\bibnamefont {Kutin}}, \bibinfo {author} {\bibfnamefont
  {N.}~\bibnamefont {Linden}}, \bibinfo {author} {\bibfnamefont
  {D.}~\bibnamefont {Shepherd}}, \ and\ \bibinfo {author} {\bibfnamefont
  {M.}~\bibnamefont {Stather}},\ }\href {\doibase 10.1098/rspa.2012.0686}
  {\bibfield  {journal} {\bibinfo  {journal} {Proceedings of the Royal Society
  of London A: Mathematical, Physical and Engineering Sciences}\ }\textbf
  {\bibinfo {volume} {469}},\ \bibinfo {pages} {1} (\bibinfo {year}
  {2013})}\BibitemShut {NoStop}%
\bibitem [{\citenamefont {Bennett}\ \emph {et~al.}(1993)\citenamefont
  {Bennett}, \citenamefont {Crepeau}, \citenamefont {Jozsa},\ and\
  \citenamefont {Peres}}]{Bennett1993TeleportingChannels}%
  \BibitemOpen
  \bibfield  {author} {\bibinfo {author} {\bibfnamefont {C.~H.}\ \bibnamefont
  {Bennett}}, \bibinfo {author} {\bibfnamefont {C.}~\bibnamefont {Crepeau}},
  \bibinfo {author} {\bibfnamefont {R.}~\bibnamefont {Jozsa}}, \ and\ \bibinfo
  {author} {\bibfnamefont {A.}~\bibnamefont {Peres}},\ }\href@noop {}
  {\bibfield  {journal} {\bibinfo  {journal} {Physical Review Letters}\
  }\textbf {\bibinfo {volume} {70}} (\bibinfo {year} {1993})}\BibitemShut
  {NoStop}%
\bibitem [{\citenamefont {Zwerger}\ \emph {et~al.}(2012)\citenamefont
  {Zwerger}, \citenamefont {D{\"{u}}r},\ and\ \citenamefont
  {Briegel}}]{Zwerger2012Measurement-basedRepeaters}%
  \BibitemOpen
  \bibfield  {author} {\bibinfo {author} {\bibfnamefont {M.}~\bibnamefont
  {Zwerger}}, \bibinfo {author} {\bibfnamefont {W.}~\bibnamefont {D{\"{u}}r}},
  \ and\ \bibinfo {author} {\bibfnamefont {H.~J.}\ \bibnamefont {Briegel}},\
  }\href {\doibase 10.1103/PhysRevA.85.062326} {\bibfield  {journal} {\bibinfo
  {journal} {Physical Review A}\ }\textbf {\bibinfo {volume} {85}},\ \bibinfo
  {pages} {1} (\bibinfo {year} {2012})}\BibitemShut {NoStop}%
\bibitem [{\citenamefont {Azuma}\ \emph {et~al.}(2015)\citenamefont {Azuma},
  \citenamefont {Tamaki},\ and\ \citenamefont
  {Lo}}]{Azuma2015All-photonicRepeaters}%
  \BibitemOpen
  \bibfield  {author} {\bibinfo {author} {\bibfnamefont {K.}~\bibnamefont
  {Azuma}}, \bibinfo {author} {\bibfnamefont {K.}~\bibnamefont {Tamaki}}, \
  and\ \bibinfo {author} {\bibfnamefont {H.-K.}\ \bibnamefont {Lo}},\ }\href
  {\doibase 10.1038/ncomms7787} {\bibfield  {journal} {\bibinfo  {journal}
  {Nature Communications}\ }\textbf {\bibinfo {volume} {6}},\ \bibinfo {pages}
  {6787} (\bibinfo {year} {2015})}\BibitemShut {NoStop}%
\bibitem [{\citenamefont {Jiang}\ \emph {et~al.}(2009)\citenamefont {Jiang},
  \citenamefont {Taylor}, \citenamefont {Nemoto}, \citenamefont {Munro},
  \citenamefont {Van~Meter},\ and\ \citenamefont
  {Lukin}}]{Jiang2009QuantumEncoding}%
  \BibitemOpen
  \bibfield  {author} {\bibinfo {author} {\bibfnamefont {L.}~\bibnamefont
  {Jiang}}, \bibinfo {author} {\bibfnamefont {J.~M.}\ \bibnamefont {Taylor}},
  \bibinfo {author} {\bibfnamefont {K.}~\bibnamefont {Nemoto}}, \bibinfo
  {author} {\bibfnamefont {W.~J.}\ \bibnamefont {Munro}}, \bibinfo {author}
  {\bibfnamefont {R.}~\bibnamefont {Van~Meter}}, \ and\ \bibinfo {author}
  {\bibfnamefont {M.~D.}\ \bibnamefont {Lukin}},\ }\href {\doibase
  10.1103/PhysRevA.79.032325} {\bibfield  {journal} {\bibinfo  {journal}
  {Physical Review A}\ }\textbf {\bibinfo {volume} {79}},\ \bibinfo {pages}
  {032325} (\bibinfo {year} {2009})}\BibitemShut {NoStop}%
\bibitem [{\citenamefont {Simon}\ \emph {et~al.}(2007)\citenamefont {Simon},
  \citenamefont {De~Riedmatten}, \citenamefont {Afzelius}, \citenamefont
  {Sangouard}, \citenamefont {Zbinden},\ and\ \citenamefont
  {Gisin}}]{Simon2007QuantumMemories}%
  \BibitemOpen
  \bibfield  {author} {\bibinfo {author} {\bibfnamefont {C.}~\bibnamefont
  {Simon}}, \bibinfo {author} {\bibfnamefont {H.}~\bibnamefont
  {De~Riedmatten}}, \bibinfo {author} {\bibfnamefont {M.}~\bibnamefont
  {Afzelius}}, \bibinfo {author} {\bibfnamefont {N.}~\bibnamefont {Sangouard}},
  \bibinfo {author} {\bibfnamefont {H.}~\bibnamefont {Zbinden}}, \ and\
  \bibinfo {author} {\bibfnamefont {N.}~\bibnamefont {Gisin}},\ }\href
  {\doibase 10.1103/PhysRevLett.98.190503} {\bibfield  {journal} {\bibinfo
  {journal} {Physical Review Letters}\ }\textbf {\bibinfo {volume} {98}},\
  \bibinfo {pages} {1} (\bibinfo {year} {2007})}\BibitemShut {NoStop}%
\bibitem [{\citenamefont {Kok}\ \emph {et~al.}(2003)\citenamefont {Kok},
  \citenamefont {Williams},\ and\ \citenamefont
  {Dowling}}]{Kok2003ConstructionOptics}%
  \BibitemOpen
  \bibfield  {author} {\bibinfo {author} {\bibfnamefont {P.}~\bibnamefont
  {Kok}}, \bibinfo {author} {\bibfnamefont {C.~P.}\ \bibnamefont {Williams}}, \
  and\ \bibinfo {author} {\bibfnamefont {J.~P.}\ \bibnamefont {Dowling}},\
  }\href {\doibase 10.1103/PhysRevA.68.022301} {\bibfield  {journal} {\bibinfo
  {journal} {Physical Review A}\ }\textbf {\bibinfo {volume} {68}},\ \bibinfo
  {pages} {022301} (\bibinfo {year} {2003})}\BibitemShut {NoStop}%
\bibitem [{\citenamefont {Duan}\ \emph {et~al.}(2001)\citenamefont {Duan},
  \citenamefont {Lukin}, \citenamefont {Cirac},\ and\ \citenamefont
  {Zoller}}]{Duan2001LongdistanceOptics}%
  \BibitemOpen
  \bibfield  {author} {\bibinfo {author} {\bibfnamefont {L.~M.}\ \bibnamefont
  {Duan}}, \bibinfo {author} {\bibfnamefont {M.~D.}\ \bibnamefont {Lukin}},
  \bibinfo {author} {\bibfnamefont {J.~I.}\ \bibnamefont {Cirac}}, \ and\
  \bibinfo {author} {\bibfnamefont {P.}~\bibnamefont {Zoller}},\ }\href
  {\doibase 10.1038/35106500} {\bibfield  {journal} {\bibinfo  {journal}
  {Nature}\ }\textbf {\bibinfo {volume} {414}},\ \bibinfo {pages} {413}
  (\bibinfo {year} {2001})}\BibitemShut {NoStop}%
\bibitem [{\citenamefont {Hensen}\ \emph {et~al.}(2015)\citenamefont {Hensen},
  \citenamefont {Bernien}, \citenamefont {Dr{\'{e}}au}, \citenamefont
  {Reiserer}, \citenamefont {Kalb}, \citenamefont {Blok}, \citenamefont
  {Ruitenberg}, \citenamefont {Vermeulen}, \citenamefont {Schouten},
  \citenamefont {Abell{\'{a}}n}, \citenamefont {Amaya}, \citenamefont
  {Pruneri}, \citenamefont {Mitchell}, \citenamefont {Markham}, \citenamefont
  {Twitchen}, \citenamefont {Elkouss}, \citenamefont {Wehner}, \citenamefont
  {Taminiau},\ and\ \citenamefont
  {Hanson}}]{Hensen2015Loophole-freeKilometres}%
  \BibitemOpen
  \bibfield  {author} {\bibinfo {author} {\bibfnamefont {B.}~\bibnamefont
  {Hensen}}, \bibinfo {author} {\bibfnamefont {H.}~\bibnamefont {Bernien}},
  \bibinfo {author} {\bibfnamefont {A.~E.}\ \bibnamefont {Dr{\'{e}}au}},
  \bibinfo {author} {\bibfnamefont {A.}~\bibnamefont {Reiserer}}, \bibinfo
  {author} {\bibfnamefont {N.}~\bibnamefont {Kalb}}, \bibinfo {author}
  {\bibfnamefont {M.~S.}\ \bibnamefont {Blok}}, \bibinfo {author}
  {\bibfnamefont {J.}~\bibnamefont {Ruitenberg}}, \bibinfo {author}
  {\bibfnamefont {R.~F.~L.}\ \bibnamefont {Vermeulen}}, \bibinfo {author}
  {\bibfnamefont {R.~N.}\ \bibnamefont {Schouten}}, \bibinfo {author}
  {\bibfnamefont {C.}~\bibnamefont {Abell{\'{a}}n}}, \bibinfo {author}
  {\bibfnamefont {W.}~\bibnamefont {Amaya}}, \bibinfo {author} {\bibfnamefont
  {V.}~\bibnamefont {Pruneri}}, \bibinfo {author} {\bibfnamefont {M.~W.}\
  \bibnamefont {Mitchell}}, \bibinfo {author} {\bibfnamefont {M.}~\bibnamefont
  {Markham}}, \bibinfo {author} {\bibfnamefont {D.~J.}\ \bibnamefont
  {Twitchen}}, \bibinfo {author} {\bibfnamefont {D.}~\bibnamefont {Elkouss}},
  \bibinfo {author} {\bibfnamefont {S.}~\bibnamefont {Wehner}}, \bibinfo
  {author} {\bibfnamefont {T.~H.}\ \bibnamefont {Taminiau}}, \ and\ \bibinfo
  {author} {\bibfnamefont {R.}~\bibnamefont {Hanson}},\ }\href {\doibase
  10.1038/nature15759} {\bibfield  {journal} {\bibinfo  {journal} {Nature}\
  }\textbf {\bibinfo {volume} {526}},\ \bibinfo {pages} {682} (\bibinfo {year}
  {2015})}\BibitemShut {NoStop}%
\bibitem [{\citenamefont {Pfaff}\ \emph {et~al.}(2014)\citenamefont {Pfaff},
  \citenamefont {Hensen}, \citenamefont {Bernien}, \citenamefont {van Dam},
  \citenamefont {Blok}, \citenamefont {Taminiau}, \citenamefont {Tiggelman},
  \citenamefont {Schouten}, \citenamefont {Markham}, \citenamefont {Twitchen},\
  and\ \citenamefont {Hanson}}]{Pfaff2014UnconditionalQubits}%
  \BibitemOpen
  \bibfield  {author} {\bibinfo {author} {\bibfnamefont {W.}~\bibnamefont
  {Pfaff}}, \bibinfo {author} {\bibfnamefont {B.}~\bibnamefont {Hensen}},
  \bibinfo {author} {\bibfnamefont {H.}~\bibnamefont {Bernien}}, \bibinfo
  {author} {\bibfnamefont {S.~B.}\ \bibnamefont {van Dam}}, \bibinfo {author}
  {\bibfnamefont {M.~S.}\ \bibnamefont {Blok}}, \bibinfo {author}
  {\bibfnamefont {T.~H.}\ \bibnamefont {Taminiau}}, \bibinfo {author}
  {\bibfnamefont {M.~J.}\ \bibnamefont {Tiggelman}}, \bibinfo {author}
  {\bibfnamefont {R.~N.}\ \bibnamefont {Schouten}}, \bibinfo {author}
  {\bibfnamefont {M.}~\bibnamefont {Markham}}, \bibinfo {author} {\bibfnamefont
  {D.~J.}\ \bibnamefont {Twitchen}}, \ and\ \bibinfo {author} {\bibfnamefont
  {R.}~\bibnamefont {Hanson}},\ }\href {\doibase 10.1126/science.1253512}
  {\bibfield  {journal} {\bibinfo  {journal} {Science}\ }\textbf {\bibinfo
  {volume} {345}},\ \bibinfo {pages} {5} (\bibinfo {year} {2014})}\BibitemShut
  {NoStop}%
\bibitem [{\citenamefont {Barrett}\ and\ \citenamefont
  {Kok}(2005)}]{Barrett2005EfficientOptics}%
  \BibitemOpen
  \bibfield  {author} {\bibinfo {author} {\bibfnamefont {S.~D.}\ \bibnamefont
  {Barrett}}\ and\ \bibinfo {author} {\bibfnamefont {P.}~\bibnamefont {Kok}},\
  }\href {\doibase 10.1103/PhysRevA.71.060310} {\bibfield  {journal} {\bibinfo
  {journal} {Physical Review A}\ }\textbf {\bibinfo {volume} {71}},\ \bibinfo
  {pages} {060310} (\bibinfo {year} {2005})}\BibitemShut {NoStop}%
\bibitem [{\citenamefont {Childress}\ \emph {et~al.}(2006)\citenamefont
  {Childress}, \citenamefont {Taylor}, \citenamefont {S{\o}rensen},\ and\
  \citenamefont {Lukin}}]{Childress2006Fault-tolerantEmitters}%
  \BibitemOpen
  \bibfield  {author} {\bibinfo {author} {\bibfnamefont {L.}~\bibnamefont
  {Childress}}, \bibinfo {author} {\bibfnamefont {J.~M.}\ \bibnamefont
  {Taylor}}, \bibinfo {author} {\bibfnamefont {A.~S.}\ \bibnamefont
  {S{\o}rensen}}, \ and\ \bibinfo {author} {\bibfnamefont {M.~D.}\ \bibnamefont
  {Lukin}},\ }\href {\doibase 10.1103/PhysRevLett.96.070504} {\bibfield
  {journal} {\bibinfo  {journal} {Physical Review Letters}\ }\textbf {\bibinfo
  {volume} {96}},\ \bibinfo {pages} {96} (\bibinfo {year} {2006})}\BibitemShut
  {NoStop}%
\bibitem [{\citenamefont {Nemoto}\ \emph {et~al.}(2016)\citenamefont {Nemoto},
  \citenamefont {Trupke}, \citenamefont {Devitt}, \citenamefont
  {Scharfenberger}, \citenamefont {Buczak}, \citenamefont {Schmiedmayer},\ and\
  \citenamefont {Munro}}]{Nemoto2014PhotonicCenters}%
  \BibitemOpen
  \bibfield  {author} {\bibinfo {author} {\bibfnamefont {K.}~\bibnamefont
  {Nemoto}}, \bibinfo {author} {\bibfnamefont {M.}~\bibnamefont {Trupke}},
  \bibinfo {author} {\bibfnamefont {S.~J.}\ \bibnamefont {Devitt}}, \bibinfo
  {author} {\bibfnamefont {B.}~\bibnamefont {Scharfenberger}}, \bibinfo
  {author} {\bibfnamefont {K.}~\bibnamefont {Buczak}}, \bibinfo {author}
  {\bibfnamefont {J.}~\bibnamefont {Schmiedmayer}}, \ and\ \bibinfo {author}
  {\bibfnamefont {W.~J.}\ \bibnamefont {Munro}},\ }\href@noop {} {\bibfield
  {journal} {\bibinfo  {journal} {Scientific reports}\ }\textbf {\bibinfo
  {volume} {6}} (\bibinfo {year} {2016})}\BibitemShut {NoStop}%
\bibitem [{\citenamefont {Jelezko}\ \emph {et~al.}(2004)\citenamefont
  {Jelezko}, \citenamefont {Gaebel}, \citenamefont {Popa}, \citenamefont
  {Gruber},\ and\ \citenamefont {Wrachtrup}}]{Jelezko2004ObservationSpin}%
  \BibitemOpen
  \bibfield  {author} {\bibinfo {author} {\bibfnamefont {F.}~\bibnamefont
  {Jelezko}}, \bibinfo {author} {\bibfnamefont {T.}~\bibnamefont {Gaebel}},
  \bibinfo {author} {\bibfnamefont {I.}~\bibnamefont {Popa}}, \bibinfo {author}
  {\bibfnamefont {A.}~\bibnamefont {Gruber}}, \ and\ \bibinfo {author}
  {\bibfnamefont {J.}~\bibnamefont {Wrachtrup}},\ }\href {\doibase
  10.1103/PhysRevLett.92.076401} {\bibfield  {journal} {\bibinfo  {journal}
  {Physical Review Letters}\ }\textbf {\bibinfo {volume} {92}},\ \bibinfo
  {pages} {076401} (\bibinfo {year} {2004})}\BibitemShut {NoStop}%
\bibitem [{\citenamefont {Ladd}\ \emph {et~al.}(2005)\citenamefont {Ladd},
  \citenamefont {Maryenko}, \citenamefont {Yamamoto}, \citenamefont {Abe},\
  and\ \citenamefont {Itoh}}]{Ladd2005CoherenceSilicon}%
  \BibitemOpen
  \bibfield  {author} {\bibinfo {author} {\bibfnamefont {T.~D.}\ \bibnamefont
  {Ladd}}, \bibinfo {author} {\bibfnamefont {D.}~\bibnamefont {Maryenko}},
  \bibinfo {author} {\bibfnamefont {Y.}~\bibnamefont {Yamamoto}}, \bibinfo
  {author} {\bibfnamefont {E.}~\bibnamefont {Abe}}, \ and\ \bibinfo {author}
  {\bibfnamefont {K.~M.}\ \bibnamefont {Itoh}},\ }\href {\doibase
  10.1103/PhysRevB.71.014401} {\bibfield  {journal} {\bibinfo  {journal}
  {Physical Review B}\ }\textbf {\bibinfo {volume} {71}},\ \bibinfo {pages} {1}
  (\bibinfo {year} {2005})}\BibitemShut {NoStop}%
\bibitem [{\citenamefont {Benjamin}\ \emph {et~al.}(2006)\citenamefont
  {Benjamin}, \citenamefont {Browne}, \citenamefont {Fitzsimons},\ and\
  \citenamefont {Morton}}]{Benjamin2006BrokeredComputation}%
  \BibitemOpen
  \bibfield  {author} {\bibinfo {author} {\bibfnamefont {S.~C.}\ \bibnamefont
  {Benjamin}}, \bibinfo {author} {\bibfnamefont {D.~E.}\ \bibnamefont
  {Browne}}, \bibinfo {author} {\bibfnamefont {J.}~\bibnamefont {Fitzsimons}},
  \ and\ \bibinfo {author} {\bibfnamefont {J.~J.~L.}\ \bibnamefont {Morton}},\
  }\href@noop {} {\bibfield  {journal} {\bibinfo  {journal} {New Journal of
  Physics}\ }\textbf {\bibinfo {volume} {8}} (\bibinfo {year}
  {2006})}\BibitemShut {NoStop}%
\bibitem [{\citenamefont {Deutsch}\ \emph {et~al.}(1996)\citenamefont
  {Deutsch}, \citenamefont {Ekert}, \citenamefont {Jozsa}, \citenamefont
  {Macchiavello}, \citenamefont {Popescu},\ and\ \citenamefont
  {Sanpera}}]{Deutsch1996QuantumChannels}%
  \BibitemOpen
  \bibfield  {author} {\bibinfo {author} {\bibfnamefont {D.}~\bibnamefont
  {Deutsch}}, \bibinfo {author} {\bibfnamefont {A.}~\bibnamefont {Ekert}},
  \bibinfo {author} {\bibfnamefont {R.}~\bibnamefont {Jozsa}}, \bibinfo
  {author} {\bibfnamefont {C.}~\bibnamefont {Macchiavello}}, \bibinfo {author}
  {\bibfnamefont {S.}~\bibnamefont {Popescu}}, \ and\ \bibinfo {author}
  {\bibfnamefont {A.}~\bibnamefont {Sanpera}},\ }\href {\doibase
  10.1103/PhysRevLett.77.2818} {\bibfield  {journal} {\bibinfo  {journal}
  {Physical Review Letters}\ }\textbf {\bibinfo {volume} {77}},\ \bibinfo
  {pages} {2818} (\bibinfo {year} {1996})}\BibitemShut {NoStop}%
\bibitem [{\citenamefont {Li}\ \emph {et~al.}(2013)\citenamefont {Li},
  \citenamefont {Barrett}, \citenamefont {Stace},\ and\ \citenamefont
  {Benjamin}}]{Li2013LongDistribution}%
  \BibitemOpen
  \bibfield  {author} {\bibinfo {author} {\bibfnamefont {Y.}~\bibnamefont
  {Li}}, \bibinfo {author} {\bibfnamefont {S.~D.}\ \bibnamefont {Barrett}},
  \bibinfo {author} {\bibfnamefont {T.~M.}\ \bibnamefont {Stace}}, \ and\
  \bibinfo {author} {\bibfnamefont {S.~C.}\ \bibnamefont {Benjamin}},\ }\href
  {\doibase 10.1088/1367-2630/15/2/023012} {\bibfield  {journal} {\bibinfo
  {journal} {New Journal of Physics}\ }\textbf {\bibinfo {volume} {15}},\
  \bibinfo {pages} {023012} (\bibinfo {year} {2013})}\BibitemShut {NoStop}%
\bibitem [{\citenamefont {Epping}\ \emph {et~al.}(2015)\citenamefont {Epping},
  \citenamefont {Kampermann},\ and\ \citenamefont
  {Bru{\ss}}}]{Epping2015GraphNetworks}%
  \BibitemOpen
  \bibfield  {author} {\bibinfo {author} {\bibfnamefont {M.}~\bibnamefont
  {Epping}}, \bibinfo {author} {\bibfnamefont {H.}~\bibnamefont {Kampermann}},
  \ and\ \bibinfo {author} {\bibfnamefont {D.}~\bibnamefont {Bru{\ss}}},\
  }\href@noop {} {\bibfield  {journal} {\bibinfo  {journal} {arXiv}\ }
  (\bibinfo {year} {2015})}\BibitemShut {NoStop}%
\bibitem [{\citenamefont {Hein}\ \emph {et~al.}(2004)\citenamefont {Hein},
  \citenamefont {Eisert},\ and\ \citenamefont
  {Briegel}}]{Hein2004MultipartyStates}%
  \BibitemOpen
  \bibfield  {author} {\bibinfo {author} {\bibfnamefont {M.}~\bibnamefont
  {Hein}}, \bibinfo {author} {\bibfnamefont {J.}~\bibnamefont {Eisert}}, \ and\
  \bibinfo {author} {\bibfnamefont {H.~J.}\ \bibnamefont {Briegel}},\ }\href
  {\doibase 10.1103/PhysRevA.69.062311} {\bibfield  {journal} {\bibinfo
  {journal} {Physical Review A}\ }\textbf {\bibinfo {volume} {69}},\ \bibinfo
  {pages} {062311} (\bibinfo {year} {2004})}\BibitemShut {NoStop}%
\bibitem [{\citenamefont {Gottesman}\ \emph {et~al.}(2004)\citenamefont
  {Gottesman}, \citenamefont {Lo}, \citenamefont {Lutkenhaus},\ and\
  \citenamefont {Preskill}}]{Gottesman2004SecurityDevices}%
  \BibitemOpen
  \bibfield  {author} {\bibinfo {author} {\bibfnamefont {D.}~\bibnamefont
  {Gottesman}}, \bibinfo {author} {\bibfnamefont {H.-K.}\ \bibnamefont {Lo}},
  \bibinfo {author} {\bibfnamefont {N.}~\bibnamefont {Lutkenhaus}}, \ and\
  \bibinfo {author} {\bibfnamefont {J.}~\bibnamefont {Preskill}},\ }in\
  \href@noop {} {\emph {\bibinfo {booktitle} {Information Theory, 2004. ISIT
  2004. Proceedings. International Symposium on}}}\ (\bibinfo {organization}
  {IEEE},\ \bibinfo {year} {2004})\ p.\ \bibinfo {pages} {136}\BibitemShut
  {NoStop}%
\bibitem [{\citenamefont {Renner}(2008)}]{Renner2005SecurityDistribution}%
  \BibitemOpen
  \bibfield  {author} {\bibinfo {author} {\bibfnamefont {R.}~\bibnamefont
  {Renner}},\ }\href@noop {} {\bibfield  {journal} {\bibinfo  {journal}
  {International Journal of Quantum Information}\ }\textbf {\bibinfo {volume}
  {6}},\ \bibinfo {pages} {1} (\bibinfo {year} {2008})}\BibitemShut {NoStop}%
\bibitem [{\citenamefont {Englund}\ \emph {et~al.}(2010)\citenamefont
  {Englund}, \citenamefont {Shields}, \citenamefont {Rivoire}, \citenamefont
  {Hatami}, \citenamefont {Vuckovic}, \citenamefont {Park},\ and\ \citenamefont
  {Lukin}}]{englund2010deterministic}%
  \BibitemOpen
  \bibfield  {author} {\bibinfo {author} {\bibfnamefont {D.}~\bibnamefont
  {Englund}}, \bibinfo {author} {\bibfnamefont {B.}~\bibnamefont {Shields}},
  \bibinfo {author} {\bibfnamefont {K.}~\bibnamefont {Rivoire}}, \bibinfo
  {author} {\bibfnamefont {F.}~\bibnamefont {Hatami}}, \bibinfo {author}
  {\bibfnamefont {J.}~\bibnamefont {Vuckovic}}, \bibinfo {author}
  {\bibfnamefont {H.}~\bibnamefont {Park}}, \ and\ \bibinfo {author}
  {\bibfnamefont {M.~D.}\ \bibnamefont {Lukin}},\ }\href@noop {} {\bibfield
  {journal} {\bibinfo  {journal} {Nano letters}\ }\textbf {\bibinfo {volume}
  {10}},\ \bibinfo {pages} {3922} (\bibinfo {year} {2010})}\BibitemShut
  {NoStop}%
\bibitem [{\citenamefont {Johnson}\ \emph {et~al.}(2015)\citenamefont
  {Johnson}, \citenamefont {Dolan}, \citenamefont {Grange}, \citenamefont
  {Trichet}, \citenamefont {Hornecker}, \citenamefont {Chen}, \citenamefont
  {Weng}, \citenamefont {Hughes}, \citenamefont {Watt}, \citenamefont
  {Auff{\`e}ves} \emph {et~al.}}]{johnson2015tunable}%
  \BibitemOpen
  \bibfield  {author} {\bibinfo {author} {\bibfnamefont {S.}~\bibnamefont
  {Johnson}}, \bibinfo {author} {\bibfnamefont {P.}~\bibnamefont {Dolan}},
  \bibinfo {author} {\bibfnamefont {T.}~\bibnamefont {Grange}}, \bibinfo
  {author} {\bibfnamefont {A.}~\bibnamefont {Trichet}}, \bibinfo {author}
  {\bibfnamefont {G.}~\bibnamefont {Hornecker}}, \bibinfo {author}
  {\bibfnamefont {Y.-C.}\ \bibnamefont {Chen}}, \bibinfo {author}
  {\bibfnamefont {L.}~\bibnamefont {Weng}}, \bibinfo {author} {\bibfnamefont
  {G.}~\bibnamefont {Hughes}}, \bibinfo {author} {\bibfnamefont
  {A.}~\bibnamefont {Watt}}, \bibinfo {author} {\bibfnamefont {A.}~\bibnamefont
  {Auff{\`e}ves}},  \emph {et~al.},\ }\href@noop {} {\bibfield  {journal}
  {\bibinfo  {journal} {New Journal of Physics}\ }\textbf {\bibinfo {volume}
  {17}},\ \bibinfo {pages} {122003} (\bibinfo {year} {2015})}\BibitemShut
  {NoStop}%
\bibitem [{\citenamefont {Leibfried}\ \emph {et~al.}(2003)\citenamefont
  {Leibfried}, \citenamefont {DeMarco}, \citenamefont {Meyer}, \citenamefont
  {Lucas}, \citenamefont {Barrett}, \citenamefont {Britton}, \citenamefont
  {Itano}, \citenamefont {Jelenkovi{\'{c}}}, \citenamefont {Langer},
  \citenamefont {Rosenband},\ and\ \citenamefont
  {Wineland}}]{Leibfried2003ExperimentalGate}%
  \BibitemOpen
  \bibfield  {author} {\bibinfo {author} {\bibfnamefont {D.}~\bibnamefont
  {Leibfried}}, \bibinfo {author} {\bibfnamefont {B.}~\bibnamefont {DeMarco}},
  \bibinfo {author} {\bibfnamefont {V.}~\bibnamefont {Meyer}}, \bibinfo
  {author} {\bibfnamefont {D.}~\bibnamefont {Lucas}}, \bibinfo {author}
  {\bibfnamefont {M.}~\bibnamefont {Barrett}}, \bibinfo {author} {\bibfnamefont
  {J.}~\bibnamefont {Britton}}, \bibinfo {author} {\bibfnamefont {W.~M.}\
  \bibnamefont {Itano}}, \bibinfo {author} {\bibfnamefont {B.}~\bibnamefont
  {Jelenkovi{\'{c}}}}, \bibinfo {author} {\bibfnamefont {C.}~\bibnamefont
  {Langer}}, \bibinfo {author} {\bibfnamefont {T.}~\bibnamefont {Rosenband}}, \
  and\ \bibinfo {author} {\bibfnamefont {D.~J.}\ \bibnamefont {Wineland}},\
  }\href {\doibase 10.1038/nature01492} {\bibfield  {journal} {\bibinfo
  {journal} {Nature}\ }\textbf {\bibinfo {volume} {422}},\ \bibinfo {pages}
  {412} (\bibinfo {year} {2003})}\BibitemShut {NoStop}%
\bibitem [{\citenamefont {Robledo}\ \emph {et~al.}(2011)\citenamefont
  {Robledo}, \citenamefont {Childress}, \citenamefont {Bernien}, \citenamefont
  {Hensen}, \citenamefont {Alkemade},\ and\ \citenamefont
  {Hanson}}]{Robledo2011HighfidelityRegister}%
  \BibitemOpen
  \bibfield  {author} {\bibinfo {author} {\bibfnamefont {L.}~\bibnamefont
  {Robledo}}, \bibinfo {author} {\bibfnamefont {L.}~\bibnamefont {Childress}},
  \bibinfo {author} {\bibfnamefont {H.}~\bibnamefont {Bernien}}, \bibinfo
  {author} {\bibfnamefont {B.}~\bibnamefont {Hensen}}, \bibinfo {author}
  {\bibfnamefont {P.~F.}\ \bibnamefont {Alkemade}}, \ and\ \bibinfo {author}
  {\bibfnamefont {R.}~\bibnamefont {Hanson}},\ }\href@noop {} {\bibfield
  {journal} {\bibinfo  {journal} {Nature}\ }\textbf {\bibinfo {volume} {477}},\
  \bibinfo {pages} {574} (\bibinfo {year} {2011})}\BibitemShut {NoStop}%
\bibitem [{\citenamefont {D{\"{u}}r}\ \emph {et~al.}(1999)\citenamefont
  {D{\"{u}}r}, \citenamefont {Briegel}, \citenamefont {Cirac},\ and\
  \citenamefont {Zoller}}]{Dur1999QuantumPurification}%
  \BibitemOpen
  \bibfield  {author} {\bibinfo {author} {\bibfnamefont {W.}~\bibnamefont
  {D{\"{u}}r}}, \bibinfo {author} {\bibfnamefont {H.-J.}\ \bibnamefont
  {Briegel}}, \bibinfo {author} {\bibfnamefont {J.~I.}\ \bibnamefont {Cirac}},
  \ and\ \bibinfo {author} {\bibfnamefont {P.}~\bibnamefont {Zoller}},\ }\href
  {\doibase 10.1103/PhysRevA.59.169} {\bibfield  {journal} {\bibinfo  {journal}
  {Physical Review A}\ }\textbf {\bibinfo {volume} {59}},\ \bibinfo {pages}
  {169} (\bibinfo {year} {1999})}\BibitemShut {NoStop}%
\bibitem [{\citenamefont {Krovi}\ \emph {et~al.}(2015)\citenamefont {Krovi},
  \citenamefont {Guha}, \citenamefont {Dutton}, \citenamefont {Slater},
  \citenamefont {Simon},\ and\ \citenamefont
  {Tittel}}]{Krovi2015PracticalSources}%
  \BibitemOpen
  \bibfield  {author} {\bibinfo {author} {\bibfnamefont {H.}~\bibnamefont
  {Krovi}}, \bibinfo {author} {\bibfnamefont {S.}~\bibnamefont {Guha}},
  \bibinfo {author} {\bibfnamefont {Z.}~\bibnamefont {Dutton}}, \bibinfo
  {author} {\bibfnamefont {J.~A.}\ \bibnamefont {Slater}}, \bibinfo {author}
  {\bibfnamefont {C.}~\bibnamefont {Simon}}, \ and\ \bibinfo {author}
  {\bibfnamefont {W.}~\bibnamefont {Tittel}},\ }\href
  {http://arxiv.org/abs/1505.03470} {\ ,\ \bibinfo {pages} {19} (\bibinfo
  {year} {2015})}\BibitemShut {NoStop}%
\bibitem [{\citenamefont {Piparo}\ and\ \citenamefont
  {Razavi}(2015)}]{piparo2015long}%
  \BibitemOpen
  \bibfield  {author} {\bibinfo {author} {\bibfnamefont {N.~L.}\ \bibnamefont
  {Piparo}}\ and\ \bibinfo {author} {\bibfnamefont {M.}~\bibnamefont
  {Razavi}},\ }\href@noop {} {\bibfield  {journal} {\bibinfo  {journal} {IEEE
  Journal of Selected Topics in Quantum Electronics}\ }\textbf {\bibinfo
  {volume} {21}},\ \bibinfo {pages} {123} (\bibinfo {year} {2015})}\BibitemShut
  {NoStop}%
\bibitem [{\citenamefont {Munro}\ \emph {et~al.}(2012)\citenamefont {Munro},
  \citenamefont {Stephens}, \citenamefont {Devitt}, \citenamefont {Harrison},\
  and\ \citenamefont {Nemoto}}]{Munro2012QuantumMemories}%
  \BibitemOpen
  \bibfield  {author} {\bibinfo {author} {\bibfnamefont {W.~J.}\ \bibnamefont
  {Munro}}, \bibinfo {author} {\bibfnamefont {a.~M.}\ \bibnamefont {Stephens}},
  \bibinfo {author} {\bibfnamefont {S.~J.}\ \bibnamefont {Devitt}}, \bibinfo
  {author} {\bibfnamefont {K.~a.}\ \bibnamefont {Harrison}}, \ and\ \bibinfo
  {author} {\bibfnamefont {K.}~\bibnamefont {Nemoto}},\ }\href {\doibase
  10.1038/nphoton.2012.243} {\bibfield  {journal} {\bibinfo  {journal} {Nature
  Photonics}\ }\textbf {\bibinfo {volume} {6}},\ \bibinfo {pages} {777}
  (\bibinfo {year} {2012})}\BibitemShut {NoStop}%
\end{thebibliography}

%

\end{document}